\documentclass{article}

\usepackage{arxiv}

\usepackage[utf8]{inputenc} 
\usepackage[T1]{fontenc}    
\usepackage{hyperref}       
\usepackage{url}            
\usepackage{booktabs}       
\usepackage{amsfonts}       
\usepackage{nicefrac}       
\usepackage{microtype}      
\usepackage{lipsum}
\usepackage{cite}
\usepackage{amsmath,amssymb,amsfonts}
\usepackage{algorithmic}
\usepackage{graphicx}
\usepackage{textcomp}
\usepackage{xcolor}
\usepackage{listings}

\title{Towards Tracking Data Flows in Cloud Architectures}

\author{
  Immanuel Kunz*, Valentina Casola**, Angelika Schneider*, Christian Banse* and Julian Schütte*\\
  * Fraunhofer AISEC, Garching b. M\"unchen, Germany\\ 
  Email: {\{firstname.lastname\}@aisec.fraunhofer.de}\\
  ** University of Naples Federico II, Naples, Italy\\
  Email: casolav@unina.it
}

\begin{document}
\maketitle

\begin{abstract}
As cloud services become central in an increasing number of applications, they process and store more personal and business-critical data. At the same time, privacy and compliance regulations such as GDPR, the EU ePrivacy regulation, PCI, and the upcoming EU Cybersecurity Act raise the bar for secure processing and traceability of critical data. 
Especially the demand to provide information about existing data records of an individual and the ability to delete them on demand is central in privacy regulations. 
Common to these requirements is that cloud providers must be able to track data as it flows across the different services to ensure that it never moves outside of the legitimate realm, and it is known at all times where a specific copy of a record that belongs to a specific individual or business process is located.
However, current cloud architectures do neither provide the means to holistically track data flows across different services nor to enforce policies on data flows. In this paper, we point out the deficits in the data flow tracking functionalities of major cloud providers by means of a set of practical experiments. We then generalize from these experiments introducing a generic architecture that aims at solving the problem of cloud-wide data flow tracking and show how it can be built in a Kubernetes-based prototype implementation.
\end{abstract}

\keywords{Data provenance \and information flow control \and GDPR Compliance}

\section{Introduction}
The usage of cloud computing can bring a lot of advantages to its users, e.g. flexible, pay-as-you-go storage solutions. Yet, when storing data in the cloud, sensitive and personal data needs to be tracked, e.g. regarding its replications and geolocations. Since the General Data Protection Regulation (GDPR) became enforceable in the European Union, processing and storing personal data in the cloud has become a liability for service providers in the cloud.

Keeping track of, and enforcing data flows in cloud systems is an essential requirement for practical reasons, as well as for legal and compliance reasons, e.g. when personal data is processed and stored.
First, the GDPR establishes various requirements towards the processing of personal data. It requires data controllers to limit the storage of personal data (Art. 5) and to be able to demonstrate how they process personal data. Furthermore, it requires data controllers to provide certain rights to data subjects, such as the right to access, rectify and delete their personal data. Regarding the storage geolocation of personal data, it restricts the transfer of personal data to countries outside of the EU (Art. 44). The EU ePrivacy regulation, the upcoming certifications based on the EU Cybersecurity Act and other regulations establish comparable standards.

Also, practical requirements play a role for data controllers. For instance, tracking data flows can support data loss prevention, and data flow policy enforcement can ensure that personal data and business secrets cannot leave a defined trust boundary, preventing data breaches. 

Cloud providers offer various services to monitor resources in the cloud, as well as to manage their life-cycle. For example, Amazon Web Services (AWS) provides the possibility to configure life-cycle rules for data objects in S3 buckets. Microsoft Azure allows to configure data replication for a storage account to another region for failover safety. Yet, these services are mostly designed for data redundancy and configuration monitoring rather than for tracking data objects with the purpose of fulfilling GDPR requirements and other compliance and practical requirements.

In this paper we demonstrate how today's cloud providers fall short when it comes to tracking data flows and enforcing data flow policies, and propose a label-based tracker system to overcome these shortcomings. 
Other approaches have proposed, e.g., to tag data in a lower level of the software stack \cite{preibusch2011information} \cite{zeldovich2006making}. In contrast, we propose to label data objects on the cloud's control plane. 
We also propose a possible integration of the proposed tracker in cloud infrastructures building on existing permission systems.

The paper's contribution is threefold:

\begin{itemize}
  \item We present various data tracking mechanisms implemented in AWS, as well as compare them to other cloud providers, 
  \item we propose a label-based data tracking mechanism and a data flow-sensitive cloud architecture that can be easily deployed with different cloud providers,
  \item we present a prototypical implementation of this architecture using the Istio service mesh in Kubernetes and evaluate it in different scenarios.
\end{itemize}

The remainder of the paper is structured as follows. Section \ref{approach} describes our approach to solving specific tracking problems in cloud architectures. Hereafter, in Section \ref{architecture} we generalize from these specific solutions to a general policy enforcement architecture. Section \ref{implementation} supports this architecture with a prototypical implementation. In Section \ref{remarks}, we conclude our experimental work with some final discussions on solving data tracking problems in cloud environments. In Section \ref{relatedwork}, we describe some related work and the paper is concluded in Section \ref{conclusion}.

\section{Approach: Tracking Data in the Cloud}
\label{approach}
In this section, we present our novel approach for tracking data flows in cloud environments and, in particular, we refer to some AWS examples and also discuss the differences to Azure and Google Cloud Platform (GCP). 
We also show that comparable mechanisms, offered by cloud providers, are not sufficient to keep track of data flows and that instead, cloud users need to implement custom tracking solutions that are costly and complex. 

Consider the following example scenario which we use throughout the paper as a running example. Company \textit{CloudFlow} offers job applicants an online portal for uploading CVs and other documents. CloudFlow uses AWS to store the files in a S3 bucket. Also, they are automatically replicated to several other buckets for backup purposes. The source files and their replications, however, are only allowed to be stored in the EU. Depending on whether an applicant is rejected or accepted, her CV must be deleted after six months or can be retained. If a CV needs to be deleted, not only the source file but also all replications need to be deleted. Yet, it can be a tedious and error-prone task to search through all existing replications of a single file.

Hence, a tracking system is needed which, however, is not offered by available cloud monitoring systems. From a high level perspective, a tracking system should be able to retrieve information about data objects from a cloud service API, examine object properties (i.e metadata or custom-defined labels) and validate them against an expected policy. 

\begin{figure}[t]
\centering
\includegraphics[width=0.6\linewidth, keepaspectratio]{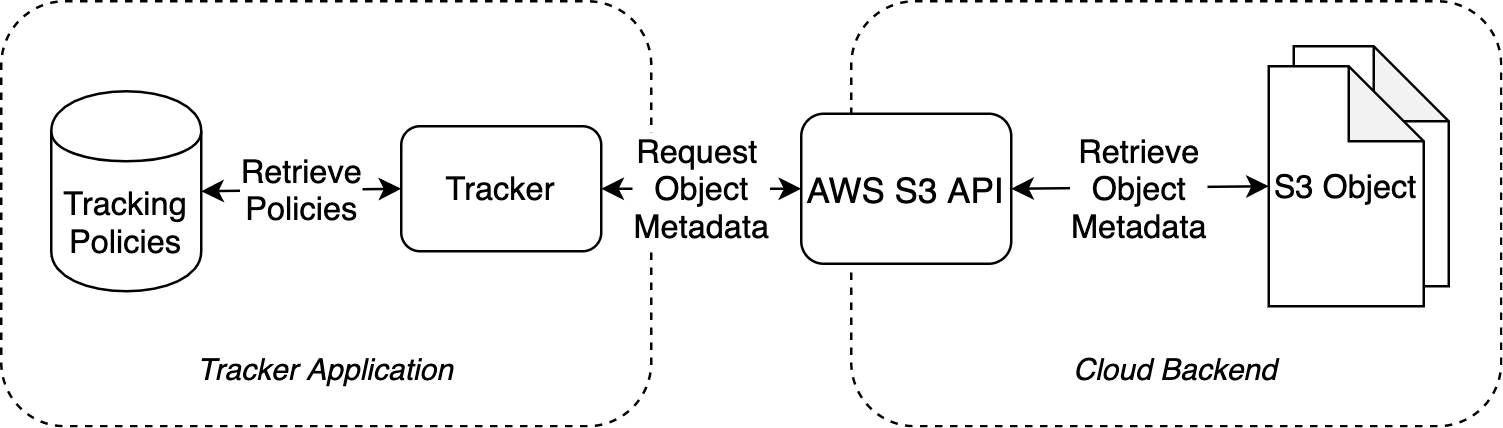}
\caption[High-level overview of the tracking system]{High-level overview of the tracking system}
\label{fig:checksarch}
\end{figure}

Figure \ref{fig:checksarch} illustrates an overview of the proposed tracking system of which the main components are the \textit{Tracker} and the \textit{Tracking Policies}.  

A tracking policy defines labels to be associated to objects and corresponding life-cycle rules. The policies are stored as a separate file. Life-cycle rules can define a retention period for a label or other metadata, as well as a deletion time. For instance, a rule may state that an object that has been labeled as \textit{[invoice]} must be deleted after one year. At the same time, it may state that data objects carrying this label must be retained for at least six months.

The Tracker first discovers existing data objects, i.e. existing storage entities and data objects are retrieved using the Cloud APIs and their titles are parsed for attached labels. Later, the labels of the discovered items are checked against the policy rules. The Tracker verifies that a data object still complies with its life-cycle policy or that the object violates it. 

The terms \textit{metadata} and \textit{label} refer to data that is associated with a data object or other resources. In the remainder of this paper we will use \textit{metadata} when we refer to already available information, as offered by AWS or other cloud providers, and we will refer to \textit{labels} to refer to custom-defined metadata which is necessary to enable some of our tracking mechanisms. The motivation for having custom-defined labels also stems from the fact that the number of \textit{tags} that can be assigned to resources natively in AWS and Azure is limited.

For example, Amazon S3 buckets metadata reveals information about the bucket's logging and permission configurations, the last modification date and whether it is a replication object or not, indicated by the \textit{REPLICA} flag. AWS allows to replicate data objects or whole S3 buckets, e.g. for data recovery purposes. When a replication for a bucket is specified, every data object that is uploaded to the bucket is replicated to the specified replication bucket. Yet, users are not notified when a replication bucket is deleted. Also, this mechanism does not delete a replication bucket when its source bucket, i.e. the bucket holding the original data objects, is deleted. This leaves the possibility that a replication bucket containing sensitive data is forgotten.

The reason that the deletion of replication buckets is not enforced in S3 is that data objects in the replication bucket may stem from more than one source bucket. Still, such orphan data objects are not tracked reliably which is why we developed a tracking mechanism which can identify such replication data objects.

In the reported scenario, desirable tracking requirements may be the following:

\begin{enumerate}
  \item Replicated data objects shall not exist without a source object.
  \item Replicated data objects shall not reside in a different geographic area than their source objects.
  \item It shall be possible to validate a data object against a custom life cycle rule, for example to check if the object should be deleted.
\end{enumerate}

In the next subsections, we describe in detail how we designed different tracking functionalities which aim at fulfilling these requirements.

\subsection{Bucket and object replication tracking}
\label{replication_tracking}
This tracking functionality aims at preventing orphan replication objects. We use the terms \textit{replication object} and \textit{replication bucket} to describe replicated data objects or replicated buckets respectively. Replications in S3 are created according to \textit{replication rules} which specify which data objects shall be replicated to which replication bucket.
As stated above, source buckets can be deleted without impacting their replications but still, deletion periods rules may apply to these replications, e.g., for compliance reasons. 
This check tracks replication objects in two ways: First, it verifies that for every replication bucket that is found, a source bucket still exists; and second, it verifies that for every replication object, its source object still exist.

To do this, it first retrieves a list of all S3 buckets and iterates through their data objects. Then, it checks whether a data object has the replication flag set to \textit{REPLICA} in its metadata. If this is the case, two separate checks are performed. 
First, it is verified that a bucket exists with a replication rule specified which points at the bucket the replication object is contained in.
While this does not guarantee that every data object's source object still exists---but only if there is any source bucket that is replicated to this replication bucket---it can still identify orphan replication buckets.
Second, it is verified that in one of the buckets whose data objects are replicated to the replication bucket, an object with the same \textit{object key}, i.e. the object's path and name uniquely identifying the object in the bucket, as the replication object exists.

A replication bucket can only be identified by iterating through its data objects and checking if there is at least one object that has the \textit{REPLICA} state set in its metadata attributes, or by identifying a replication rule that specifies it as a replication bucket. Replication rules, however, are deleted as well when a source bucket is deleted which is why it is necessary to check for replications on the object-level, too.
 

\subsection{Bucket replication geolocation tracking}
This check verifies that every replication object resides in the same region group as its source object and aims at preventing the replication to disallowed geolocations.

We define a region group as a set of regions that AWS denotes with the same two starting letters, e.g., \textit{EU} for Europe and \textit{US} for the United States. Although the goal is to verify this for every data object, we verify the objects' geolocation through the bucket they are saved in, since the information about objects' geolocation cannot be extracted from objects but only from buckets (contrary to the replication state that is saved as metadata). 

The check first retrieves a list of S3 buckets and checks for each bucket whether replication rules are specified in the tracker policies. If this is the case for a bucket, meaning that some or all of its contents are replicated to a different bucket, the specified replication buckets region group is compared to the source buckets region group. 

Notice that AWS S3 also offers a feature called \textit{Same-Region Replication} that ensures that replications can only be created in the same region as their source objects.  However, for failover reasons, it may still be important for cloud users to create replications that do not reside in the same region as the source objects, but still reside in the same region group. For GDPR compliance, for example, it may be relevant to keep the data in the EU group but still it may be reasonable to distribute replications across different regions within the EU.

\subsection{Tracking data objects' validity with custom labels} 

This check allows to validate a data object against a custom tracking policy. The validation flow is similar to the previous examples, but, in order to manage a large number of data objects and specify different custom defined rules, it is helpful to use custom-defined labels to express the semantics of different data attributes. This way, semantics are made visible on the cloud control plane where tracking or permission policies may be applied. 

In the following, we first describe this mechanism and then we present a semi-automated way to manage and associate labels.


In particular, it works in three steps as described in the following.
First, a life-cycle policy is defined which defines labels and corresponding life-cycle rules. Life-cycle rules can define a retention period for a label, as well as a deletion time. For instance, it may indicate that an object that has been associated with the label \textit{[CV]} must be deleted after one year or that data objects carrying this label must be retained for at least six months.

Then, the existing data objects are discovered, i.e. existing storage entities and data objects are retrieved using the AWS APIs and their titles are parsed for attached labels. 

In the last step, the labels of the discovered items are checked against the life-cycle configuration their labels correspond to in the custom policy. The result verifies that a data object still complies with its life-cycle policy or that the object violates it. 

In the case that the data object which is checked is a replication object, the same configuration as for its source object is applied. 

Again, considering the running example, assuming that legal regulations demand to delete CVs of rejected applicants within six months, the respective S3 buckets have to be examined regularly to ensure that CVs and their replications are deleted within their expiration time. 
Associating a common label to objects that indicates whether a file is a CV or a different file, makes it possible to build such automatic mechanisms but it could be tedious and error prone. Since there may be a very large number of objects to apply new labels to, we also propose a semi-automated mechanism which adds labels to objects based on keywords that are found in the object's title. For example, the labelling mechanisms can be configured to add the label \textit{[personal]} to all data objects that contain the keywords \textit{CV} or \textit{invoice} in their titles. The mechanism and the semantics of custom labels are important to understand the proposed tracking mechanism, and are explained in the next subsection.

\subsubsection{A semi-automated labeling mechanism} 
\label{labeling_mechanism}
The labeling mechanism first lists all data objects using the respective APIs, and then iterates through data objects' titles, scans them for pre-defined keywords and attaches them in form of a label to the title. In case no keyword is found, the user is prompted to assign a label manually. 

This mechanism is implemented using the XText framework\footnote{\url{https://www.eclipse.org/Xtext/}} which supports the development of custom programming languages.
Using XText, we defined a schema for the assignment of labels to data objects. When applying this schema, custom labels are added automatically to data objects' titles to indicate, e.g., their level of criticality or person-relation.

Listing 1 shows an example schema that defines how labels can be assigned to data objects. It is based on the running example where the company CloudFlow handles three types of data: job applications, customer data and public data.
The schema first defines the values (\textit{AssetValues}) that can be assigned to a data object, e.g. \textit{low} or \textit{personal} to indicate the level of criticality or person-relation. Furthermore, types (\textit{AssetTypes}) are defined that data objects may pertain to, i.e. the already mentioned types of data \textit{Application}, \textit{Customer} and \textit{Public}. Next, keywords (\textit{Keywords}) are listed that may be contained in the data objects' titles.
Hereafter, the schema defines rules which indicate which labels are associated to a file in case a specific keyword is found in its title. More specifically, the rules define which values (\textit{ValueIfKeyWord}) and types (\textit{TypeIfKeyword}) are assigned to a data object. In the example schema, when the keyword \textit{order} is found in the title, the labels \textit{high} and \textit{Customer} are associated to the file.
Note that all values, types and rules are completely custom-defined, different categories of labels can be easily defined and attached by simply configure the schema.
In this example, the label \textit{personal} is assigned to any data object that contains the keyword \textit{CV} in its title.

The labels assigned to data objects by such a mechanism can then be leveraged by data tracking mechanisms, e.g. to track the location of data objects more easily.

\subsection{Discussion on available cloud providers offerings}
\label{approach_discussion}
Since the previous examples were written for AWS, in this section we discuss Microsoft and Google solutions for tracking objects and how the proposed approach could be applied in Microsoft Azure and Google Cloud Providers (GCP), too.

\subsubsection{Storage and data object replication tracking}
For Azure storage accounts, replication is implemented per default. Users can only select the corresponding strategy, e.g., locally redundant storage which is stored in the same data center as the source storage. Hence, the redundancy and its deletion is managed by Azure.
Azure offers a further feature that allows to choose a specific region to which the redundant storage shall be replicated, called \textit{geo-replication}\footnote{At the time of the writing, this feature is still in preview.}. 

Also GCP's Blob storage offers similar features, like regional or multi-region redundancy. Replications of storage accounts and their locations are therefore transparent to the user. If, however, more control over data replications is required, e.g. to choose the replication region, the replication and tracking mechanisms would have to be implemented manually with custom labels and external tracking systems. 

\subsubsection{Tracking custom retention periods}
Azure allows to configure life-cycle rules for blob containers, e.g. with the effect of moving data to an archive storage or deleting the blob\footnote{Note that these life-cycle rules are different from the ones we use in the policies for our tracker system where they define retention periods}. Each option can be configured to be performed after a certain number of days after the last modification and can be applied to the whole container or certain folders inside the container.
Similar functionality is offered by GCP where life-cycle rules can be configured to delete objects or change a bucket's storage tier. 

This functionality mainly aims at reducing costs for the service users, allowing to move data to a cheaper storage or to delete it when it is not needed anymore. As in AWS S3, implementing custom retention periods is not possible. Only GCP offers the possibility to specify retention periods, e.g. object holds that prevent an object's deletion for a certain time.

In conclusion, our experiments show that AWS and Azure as well as GCP provide mechanisms for data replication for their storage services---and partly even for retention policies---but fall short when one wants to track data objects throughout their life-cycle.
In AWS S3, especially monitoring custom data retention periods is complex since data objects only provide a \textit{last modified} attribute, but not a creation date, such that buckets have to be monitored continuously for new data uploads to track their creation time. Also in Azure Blob storages, there is no efficient way of tracking replications if functionality is required that goes beyond what Azure provides by default, like distributing replications to different regions.

While services like AWS Config and Azure Policy can be used as monitoring tools (see section \ref{configpolicy}), they are designed for resource configuration monitoring rather than for data flow tracking. Implementing data tracking mechanisms nowadays is not straightforward as it requires complex customized architectures which can come at a considerable cost. In the next section, we show how the proposed tracking system can be generalized and deployed in different scenarios, including an integration into existing cloud infrastructures.

\section{High-Level Data-Flow Tracking Architecture}
\label{architecture}
In this section, we present a data-flow sensitive architecture for cloud systems which generalizes the label-based tracking mechanisms we proposed in the last section, together with different deployment models. 

\subsection{Generalized architecture}
\label{generalized_architecture}
As outlined in the previous section, the tracker system offers different functionalities and components: A discovery functionality that retrieves a list of object metadata, a validation functionality that compares an object's configuration to an expected configuration, and a policy which determines the tracking rules to be enforced.

In order to enforce tracking policies, we propose a generalized architecture for these functionalities which is inspired by modern policy enforcement architecture, as the OASIS standard XACML \cite{standard2005extensible} for attribute-based authorization decisions, which is made of an enforcement point and a decision point. 

\begin{figure}
\centering
\includegraphics[width=0.6\linewidth, keepaspectratio]{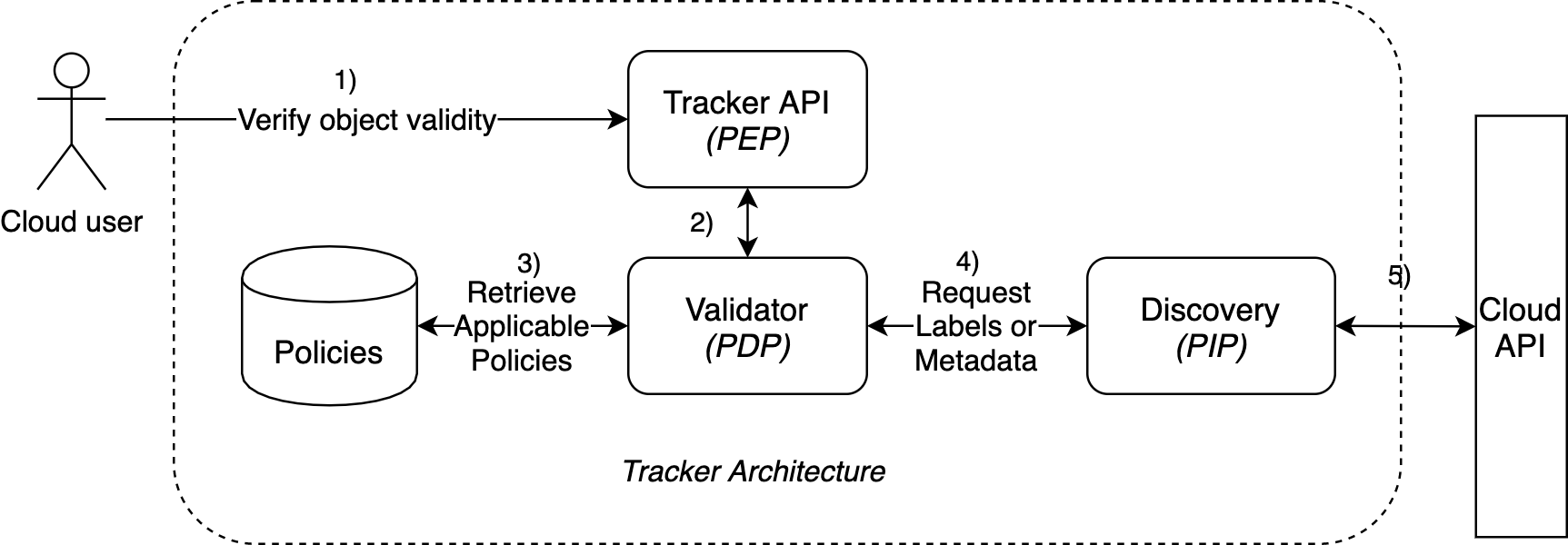}
\caption[The traditional architecture of policy enforcement systems]{The Tracker architecture and its main components}
\label{fig:traditional_pe}
\end{figure}

In Figure \ref{fig:traditional_pe}, the proposed architecture is illustrated. The Tracker System is triggered by a data flow that may originate from a user of the cloud system (1). The \textit{Tracker API} then makes a request to the \textit{Validator} (2). The Validator in turn retrieves applicable \textit{policies} (3) which specify tracking life-cycle rules.
The Validator can request the object's attribute values (metadata and labels) from the \textit{Discovery} component (4) which retrieves those values from the respective cloud APIs (5). 

Mapping this to the XACML framework, we find that the \textit{Tracker API} corresponds to a \textit{Policy Enforcement Point} (PEP), the \textit{Validator} corresponds to a \textit{Policy Decision Point} (PDP) and the \textit{Discovery} corresponds to the \textit{Policy Information Point} (PDP) as also indicated in Figure \ref{fig:traditional_pe}.

Generalizing the tracker architecture in this way allows to extend it in a modular fashion. For example, policies can be added without impacting other components and enforcement points may be triggered by different proxies and gateways provided by current technologies in their cloud control panels.

This policy enforcement approach allows to manage flows of data objects with similar properties on the control plane, rather than tracking and enforcing data flows on a lower layer of the software stack or using other mechanisms. A service provider could, for instance, decide to manage data flows by isolating resources using firewalls and resource groups. Yet, when changes need to be made, it becomes error-prone to change all corresponding firewall rules and group assignments. 

Having the cloud provider implementing an ABAC system and exposing labels on the control plane, gives the cloud users more flexibility in managing access control and simplifies the protection of data flows.
Furthermore, managing data flows based on labels complements role-based access control (RBAC) since this approach allows to separate duties of developers and roles responsible for compliance who do not need to read and understand code but only need to assign labels to resources and define rules for their circulation.

Note that for our proposed tracking system and architecture we do not impose a syntactic structure on the labels since cloud users may need to fit the labels' syntax to their specific requirements.

\subsection{Deployment models}
\label{deployments}
Given the generalized architecture, various deployment models are feasible considering different use cases for cloud users and cloud providers. 
A cloud user could set-up her own local deployment, not integrated in the cloud system, which queries external cloud APIs, similar to the high level setup described in Section \ref{approach}. 

Considering the cloud provider, we identified two possible scenarios to deploy the proposed architecture. First, a cloud provider may offer the tracking system as-a-service to complement already available services in Azure and AWS (as discussed in Section \ref{configpolicy}).

Second, this architecture can be integrated into the existing infrastructure that cloud providers already implement, mainly as Gateway APIs that can be used for security enforcement of, e.g., identity and access management (IAM) policies or RBAC authorization mechanisms.

In Azure, for example, this authorization architecture is implemented with the Azure Resource Manager (ARM). The ARM acts as a gateway API that receives requests, e.g. for creating a resource, from users and authenticates and authorizes them across services and regions.

\begin{figure}[t]
\centering
\includegraphics[width=0.5\linewidth, keepaspectratio]{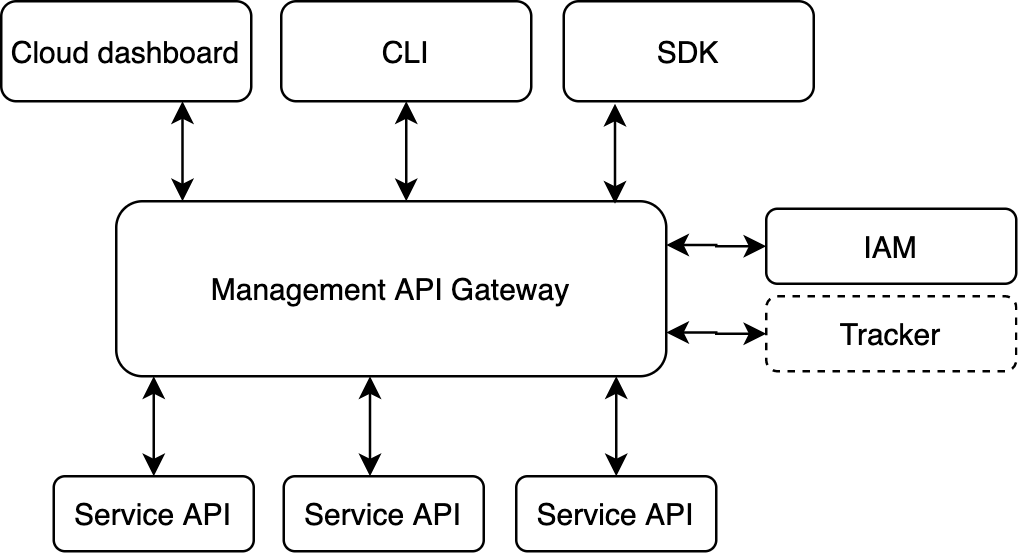}
\caption[The architecture of the central cloud API gateway]{The Tracker deployment within the central cloud API gateway}
\label{fig:label_iam}
\end{figure}

Figure \ref{fig:label_iam} illustrates this high-level architecture. It shows how clients using different kinds of mechanisms to access the cloud system, such as a dashboard interface, a CLI or a SDK, make requests to the central API gateway. This gateway performs authentication and authorization using the role-based access control (RBAC) and it is extensible. Also, the label-based tracking mechanism may be enforced here. This existing infrastructure authorizing user attributes has therefore been extended to track data attributes, defined in labels, as we will show in the next section.

\section{Prototype Implementation}
\label{implementation}
In this section, we describe our prototype implementation of the generalized tracking architecture.
We first briefly describe the implementation of the tracker system as a local client deployed by a cloud user, and its deployment in the cloud. We then describe and evaluate an implementation of the integrated deployment model as described in Section \ref{deployments} which can not only track data flows but also prevent non-compliant data flows.


For our prototype implementation we chose the integrated deployment model and set up a Kubernetes\footnote{\url{https://kubernetes.io}} cluster on virtual machines hosted in Microsoft Azure to imitate a cloud environment with multiple services which can communicate with each other, i.e. a \textit{service mesh}. We base our implementation on existing mechanisms to track and enforce policies in services meshes, such as \textit{Istio}\footnote{\url{https://istio.io}}. Istio introduces several components into a cluster including one for policy controls which is called \textit{Mixer}. 
As depicted in Figure \ref{fig:mixer_topology}, the Mixer authorizes every message flow and can be used to deduce health metrics, track data flows and enforce data flow policies. However, out-of-the-box, Istio is only able to make authorization decisions based on the metadata of the request, e.g., based on the HTTP headers the message contains or based on the attributes of source and destination on a service-level. Since our approach envisions a data flow tracking and policy enforcement based on the data's semantics (represented by labels), we implemented an extension to the Mixer, which is called an \textit{Adapter}.

While Istio itself claims to be platform-independent and several components, such as the service proxy \textit{Envoy}\footnote{\url{https://www.envoyproxy.io}} can also be used standalone, there are some benefits to the deployment into Kubernetes. For once, it allows the automatic injection of so-called \textit{sidecars} into Kubernetes Pods which act as proxies for the pods. They manage in- and outbound traffic and can relay network connections to Mixer before they are forwarded to its intended destination. As such, they can be seen as a \textit{PEP}.

Furthermore, the Mixer forwards traffic coming from the sidecar proxies according to its configuration to \textit{Adapters} which process the messages. Figure \ref{fig:mixer_topology} illustrates this showing a simplified architecture underlying the communication between services, their respective sidecar proxies and the central Mixer component. It also indicates the mapping of the Istio components to the tracker system described in Section \ref{architecture}: The service sidecars act as the \textit{Tracker API} since they enforce an authorization request to the Mixer. Furthermore, the Mixer and Adapter correspond to the \textit{Validator} since they make the decision about authorizing or rejecting a request. 
In our implementation, the Adapter additionally acts as the \textit{Discovery} component since it also retrieves a requested file's labels and makes the authorization decision on this basis. 

\subsection{Label-Based Authorization Adapter}
Using the architecture as described above, we designed a label-based tracking and policy enforcement which can not only track data objects using a monitoring approach, but also track and authorize requests as they occur. 
It is implemented as an out-of-process Istio Adapter written in Go supporting the Istio \textit{authorization} template. 
The Adapter can be configured with a list of workloads, or services, which are allowed to access data resources with a particular label.

\begin{figure}[t]
\centering
\includegraphics[width=8cm, keepaspectratio]{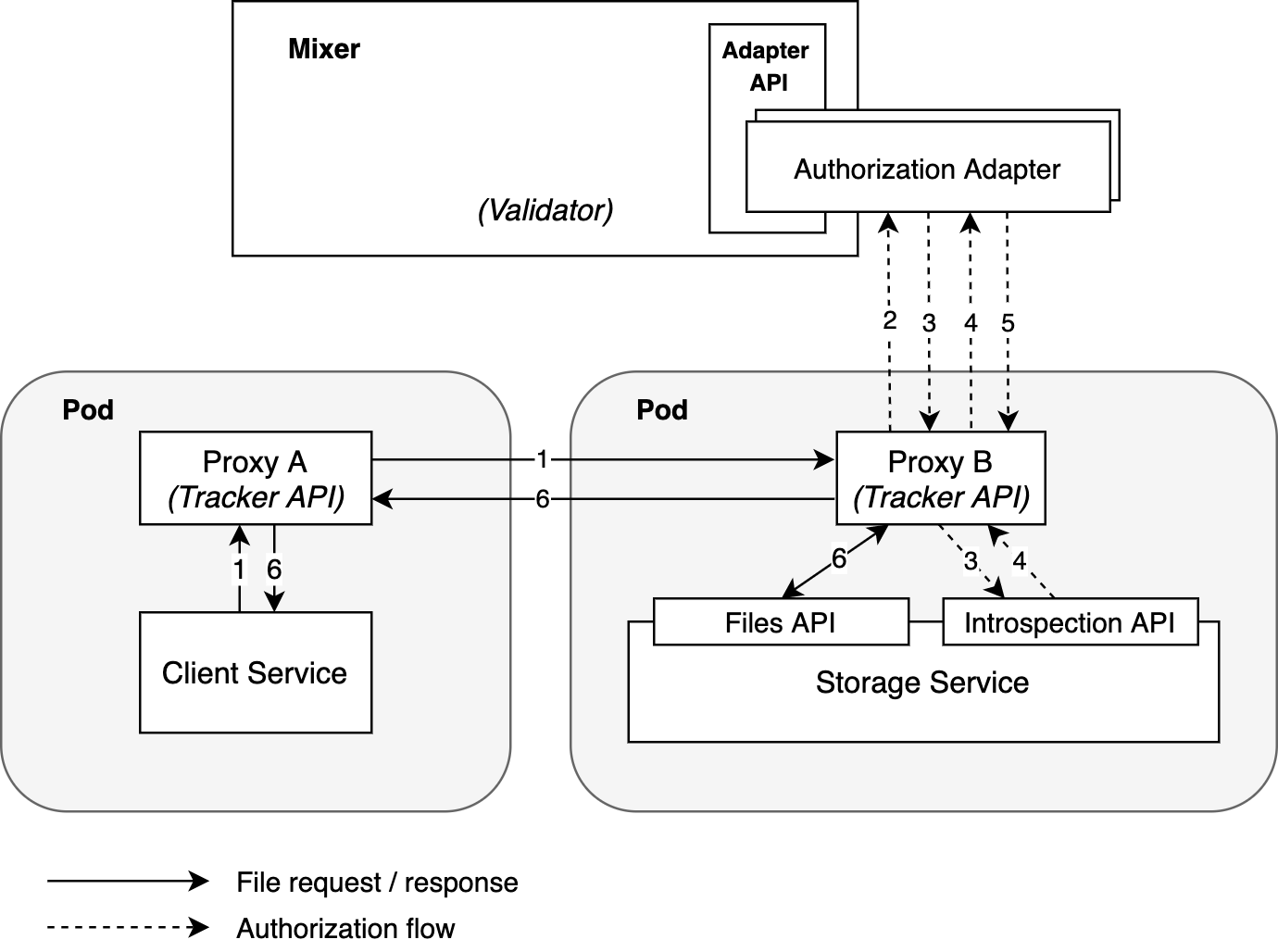}
\caption[The simplified architecture of our implementation showing an example authorization flow]{A Tracker implementation showing an example authorization flow}
\label{fig:mixer_topology}
\end{figure}

Figure \ref{fig:mixer_topology} shows an example authorization flow through this architecture: When the client service initiates a request to retrieve a file from the object storage service (OSS), it is relayed by the client's proxy to the OSS proxy (1). The OSS proxy makes an authorization request to Mixer (2) which in turn forwards the request to the Adapter. The Adapter parses the file name that is requested and inspects its labels by querying the introspection endpoint of the OSS (3).
The introspection endpoint follows a simple URL pattern of \texttt{/inspect/<resource>} and it returns the labels found at the \texttt{<resource>} using the HTTP header \texttt{X-Data-Labels} (4). 
The Adapter then compares the labels with its policy configuration which specifies the allowed workloads for certain labels. If the incoming workload is not among the list of allowed ones, the request is denied. If a resource does not contain any label, i.e. it is not classified, the request is automatically accepted. The decision is reported back to the proxy (5), and in case the request has been authorized, the proxy relays the file request to the OSS Files API and returns the response to the Client Service via its proxy (6). 

\subsection{Exemplary Service Mesh}
Our implementation uses again the running example of the company CloudFlow described in Section \ref{approach}. Part of the exemplary service mesh of CloudFlow are services for the job application portal (which handles CVs), other parts handle financial data such as invoices and public material about the company for public use. 
The mesh is configured to track and enforce the following data labels:

\begin{itemize}
  \item \textit{personal}---identifying personal data, stored in CVs. 
  \item \textit{public}---identifying non-personal information, such as a brochure about the service offering. 
  \item \textit{invoice}---indicating financial documents.
\end{itemize}

We implemented the \textit{object storage service} as a REST-based service which serves as a storage backend to store CVs and other data the company might hold. Next to the regular HTTP endpoints to retrieve files, the storage service also implements an introspection endpoint returning the labels of a specified PDF file to the Adapter. Access to the introspection endpoint is restricted to the IP address of the pod running our custom Adapter. Otherwise the labels and the existence of a file might leak to other (unauthorized) services in the cluster.

\subsection{Performance Evaluation}
\label{evaluation}
\begin{figure}[t]
\centering
\includegraphics[width=0.6\linewidth, keepaspectratio]{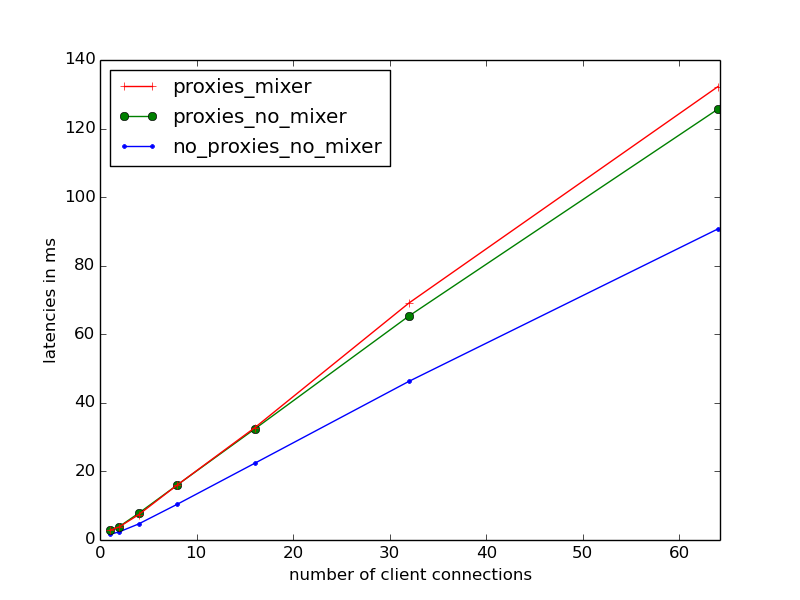}
\caption[Latencies plot against the number of client connections]{Latencies plot against the number of client connections}
\label{fig:latencies}
\end{figure}

We have compared three different scenarios to evaluate the proposed system's performance, where one client service (CS) requests files from an object storage service (OSS) as described above. The latencies plot against the number of concurrent client connections are illustrated in Figure \ref{fig:latencies}. 

In the first scenario, the two services in the Kubernetes cluster directly communicate with each other without relaying connections through proxies and enforcement of policies (\textit{no\_proxies\_no\_mixer}). 
In the second scenario, proxies are injected into the services' Pods and the request is relayed through the proxies but are not subjected to a policy enforcement by Mixer (\textit{proxies\_no\_mixer}). 
The third scenario is the one described above where the CS requests a file from the OSS via its proxy which makes an authorization request to the Mixer. The Adapter then retrieves the necessary labels from the introspection endpoint of the OSS and makes its authorization decision (\textit{proxies\_mixer}).
However, both the Envoy proxy as well as the Mixer include caches from which policy decisions can be made. This is why the latencies for the third scenario only differ slightly from the ones in the second scenario. 

To evaluate the essential overhead which is incurred by our solution, we have measured the latency that one policy check---including the call of the custom adapter to the OSS's introspection endpoint---incurs. To that end, we have set up a local installation of the Mixer server, the custom adapter, the OSS and a Mixer client. Here, one authorization request from the Mixer client was answered on average with a latency of 17.31 ms (with all caches disabled).

There can be a number of Mixer Pods in the cluster to make the policy enforcement scalable. Proxies can be granularly assigned to these Mixer instances. This scenario is also transferable to existing cloud infrastructures where the policy enforcement load may be distributed across many policy enforcement instances in different datacenters.
For further performance statistics of Istio, the interested reader is referred to the Istio benchmark tests\footnote{\url{https://istio.io/docs/ops/deployment/performance-and-scalability}}.

\section{Final remarks}
\label{remarks}
The architecture proposed in this section represents an extension to existing cloud authorization mechanisms. Such a system does not require extensive code changes or injection of monitoring monitors into virtual machines, but only acts upon custom labels defined in resources' metadata. 

Replication checks, as presented earlier, can be managed using this service by assigning the necessary labels to data objects. When they are copied, e.g. by a backup system, their labels are copied to their replications and---contrary to existing cloud providers' storage solutions---can automatically be managed in the tracking and enforcement mechanisms. 
In section \ref{approach}, we have also presented a mechanism for the tracking of data retention policies. 
Note that with the architecture proposed here, we do not aim at providing a solution for this problem. While monitoring services like Azure Policy (see Section \ref{configpolicy}) do not yet provide such a functionality, we would argue that data retention policies should be part of such monitoring services. These services monitor the \textit{state} of resources and apply actions to them according to a policy, e.g. comparing a resource's age with a retention policy. Hence, they are better suited to provide a data retention enforcement than a service focused on data \textit{flows}.

While the compliance of data geolocations could be also monitored using monitoring services as described above, this mechanism should be implemented using a data flow policy enforcement since here, a data \textit{flow} is subject to authorization. One reason is that an undesired geolocation may already present a breach of law and should be prevented rather than discovered. In our architecture, such a mechanism may be implemented by creating a corresponding label that indicates the allowed region and defining a policy on the flow of this label.

While our architecture is designed to manage data in bulk, it is possible to assign labels to data that indicate a granular relation to individual persons or devices, such that their flow across services can be logged, reconstructed and possibly subjected to data flow policies as well.

Furthermore, the proposed architecture supports the prevention of data breaches and enhances the auditability of cloud systems, since it exposes the semantics of data flows.

To manage who can attach, change, and delete labels, the existing IAM can also be extended to provide respective roles and permissions to be assigned to users. This allows not only to define which users may manage labels, but also to give resources, like virtual machines, the permission to change labels if they change data objects, e.g. by applying anonymization functions.

\section{Related Work}
\label{relatedwork}

\subsection{Information Flow Control}
Our work is related to the topic of \textit{information flow control} (IFC). In comparison to user-based access control, IFC refers to models that describe an information-centric access control, defining rules for transmissions of data and information.

Most of them also rely on labeling data in some way. For example, Pappas et al. \cite{pappas2013cloudfence} propose a framework which relies on injecting Virtual Machine Monitors on VMs' processes to tag data and track it across processes. Other works propose code annotations \cite{preibusch2011information} or customized OSs to tag and track data \cite{zeldovich2006making}.
Pasquier et al. \cite{pasquier2016data} propose an information flow control system that enforces policies using Linux Security Modules.
In \cite{suen2013s2logger}, Suen et al. propose a data tracking mechanism that tracks data at the kernel level in both host machines and virtual machines to build an end-to-end provenance control system for the cloud.
All in all, these approaches propose tracking mechanisms for lower levels of the software stack rather than building on existing high-level authorization mechanisms in cloud infrastructures.

A monitoring system directed at Platform-as-a-Service (PaaS) cloud systems called \textit{CloudSafetyNet} is proposed in \cite{priebe2014cloudsafetynet}. They propose to include so called \textit{security tags} in HTTP headers. Yet, the tags identify an application rather than expose the data's semantics to allow to manage it on the control plane.

Another work by Pasquier et al. \cite{pasquier2015expressing} proposes a cloud architecture in which a PaaS service provides audit results regarding the compliance with security policies to applications' end-users. Yet, their approach does neither target specific limitations current cloud providers have when tracking data objects, nor does it leverage the existing cloud authorization infrastructure. 

\subsection{Cloud Monitoring Services}
\label{configpolicy}
Both AWS and Azure offer services for monitoring and modifying the state of existing resources. 
AWS Config\footnote{\url{https://aws.amazon.com/config/}} provides pre-defined rules, e.g. for auditing open SSH ports in security groups, and allows to configure custom rules using the Lambda service. Here, completely customized scripts can be written in any language that Lambda supports. Hence, our experiments can also be implemented as custom AWS Config rules using standard AWS APIs. Still, Config rules can incur a significant cost if applied to a large number of resources. 

Azure Policy\footnote{\url{https://docs.microsoft.com/en-us/azure/governance/policy/overview}} offers the possibility to define policies using a specific format that includes a condition and a corresponding effect that shall be executed if the condition matches. The types of conditions and effects that can be specified, however, are limited and not customizable. At the time of the writing, it is not possible to implement a policy in the Azure Policy service that audits properties of blob objects, e.g. their creation date. Yet, if the corresponding conditions would be implemented, tracking replications, geolocations as well as custom retention periods would be implementable using such policies.

\section{Conclusion and Future Work}
\label{conclusion}
In this paper, we have presented several data tracking and validity mechanisms for data objects in cloud systems, e.g. replication-tracking of AWS S3 data objects. We have shown that today's cloud providers do not offer sufficient functionality for tracking data flows and to enforce respective policies. On this basis, we have developed a model for a data-flow sensitive cloud architecture which supports some legal data protection requirements and other practical requirements for data tracking. 

Future work includes the generalization of the tracker's rules to a reusable language that can express tracking rules.
Regarding the prototype implementation, we plan to improve the security of the label-based policy enforcement by making labels tamper-proof to protect against a malicious use of data by changing their labels.

\bibliographystyle{unsrt}
\bibliography{data_flows.bib}

\end{document}